\documentclass[sigconf]{acmart}

\usepackage{array}
\usepackage{multirow}
\usepackage{comment}
\usepackage{graphicx}
\usepackage{caption}
\usepackage{balance}      
\usepackage{graphics}     
\usepackage[T1]{fontenc}   

\usepackage{hyperref}
\usepackage{color}
\usepackage{booktabs}
\usepackage{textcomp}
\usepackage{graphics}     
\usepackage[T1]{fontenc}   

\usepackage{color}
\usepackage{booktabs}
\usepackage{textcomp}
\usepackage{comment}

\usepackage{longtable}
\usepackage{wrapfig}
\usepackage{float}
\usepackage{csquotes}
\usepackage{enumitem}
\usepackage{multirow}

\usepackage{microtype}       

\usepackage{ccicons} 

\usepackage{todonotes}

\usepackage[T1]{fontenc}
\usepackage{tabularx}
\usepackage{array}

\AtBeginDocument{
  \providecommand\BibTeX{{
    \normalfont B\kern-0.5em{\scshape i\kern-0.25em b}\kern-0.8em\TeX}}}

\copyrightyear{2022} 
\acmYear{2022} 
\setcopyright{acmcopyright}\acmConference[CHI '22 Extended Abstracts]{CHI Conference on Human Factors in Computing Systems Extended Abstracts}{April 29-May 5, 2022}{New Orleans, LA, USA}
\acmBooktitle{CHI Conference on Human Factors in Computing Systems Extended Abstracts (CHI '22 Extended Abstracts), April 29-May 5, 2022, New Orleans, LA, USA}
\acmPrice{15.00}
\acmDOI{10.1145/3491101.3519775}
\acmISBN{978-1-4503-9156-6/22/04}

\begin{document}

\title{Using Virtual Reality to Design and Evaluate a Lunar Lander: The EL3 Case Study}

\author{Tommy Nilsson} 
\affiliation{ 
    \institution{European Space Agency (ESA)}
    \city{Cologne}
    \country{Germany}} 
\email{tommy.nilsson@esa.int}

\author{Flavie Rometsch }
\affiliation{ 
    \institution{German Aerospace Center (DLR)}
    \city{Cologne}
    \country{Germany}} 
\email{flavie.rometsch@dlr.de}

\author{Andrea E. M. Casini}
\affiliation{ 
    \institution{German Aerospace Center (DLR)}
    \city{Cologne}
    \country{Germany}} 
\email{andrea.casini@esa.int}

\author{Enrico Guerra} 
\affiliation{ 
    \institution{European Space Agency (ESA)}
    \city{Cologne}
    \country{Germany}} 
\email{enricoguerra@outlook.com}

\author{Leonie Becker} 
\affiliation{ 
    \institution{European Space Agency (ESA)}
    \city{Cologne}
    \country{Germany}} 
\email{leonie2.becker@student.uni-siegen.de}

\author{Andreas Treuer}
\affiliation{ 
    \institution{European Space Agency (ESA)}
    \city{Cologne}
    \country{Germany}} 
\email{andreas.treuer@rwth-aachen.de}

\author{Paul de Medeiros}
\affiliation{ 
    \institution{European Space Agency (ESA)}
    \city{Cologne}
    \country{Germany}} 
\email{hello@pauldemedeiros.nl}

\author{Hanjo Schnellbaecher} 
\affiliation{ 
    \institution{European Space Agency (ESA)}
    \city{Cologne}
    \country{Germany}} 
\email{hanjo.schnellbaecher@ext.esa.int}

\author{Anna Vock}
\affiliation{ 
    \institution{European Space Agency (ESA)}
    \city{Cologne}
    \country{Germany}} 
\email{annalvock@gmail.com}

\author{Aidan Cowley} 
\affiliation{ 
    \institution{European Space Agency (ESA)}
    \city{Cologne}
    \country{Germany}} 
\email{aidan.cowley@esa.int}

\renewcommand{\shortauthors}{Nilsson et al.}
\renewcommand{\shorttitle}{Using Virtual Reality to Design and Evaluate a Lunar Lander: The EL3 Case Study}

\begin{abstract}
The European Large Logistics Lander (EL3) is being designed to carry out cargo delivery missions in support of future lunar ground crews. The capacity of virtual reality (VR) to visualize and interactively simulate the unique lunar environment makes it a potentially powerful design tool during the early development stages of EL3, as well as other relevant technologies. Based on input from the EL3 development team, we have produced a VR-based operational scenario featuring a hypothetical configuration of the lander. Relying on HCI research methods, we have subsequently evaluated this scenario with relevant experts (n=10). Qualitative findings from this initial pilot study have demonstrated the usefulness of VR as a design tool in this context, but likewise surfaced a number of limitations in the form of potentially impaired validity and generalizability. We conclude by outlining our future research plan and reflect on the potential use of physical stimuli to improve the validity of VR-based simulations in forthcoming design activities. 

\end{abstract}




\begin{teaserfigure}
  \includegraphics[width=\textwidth]{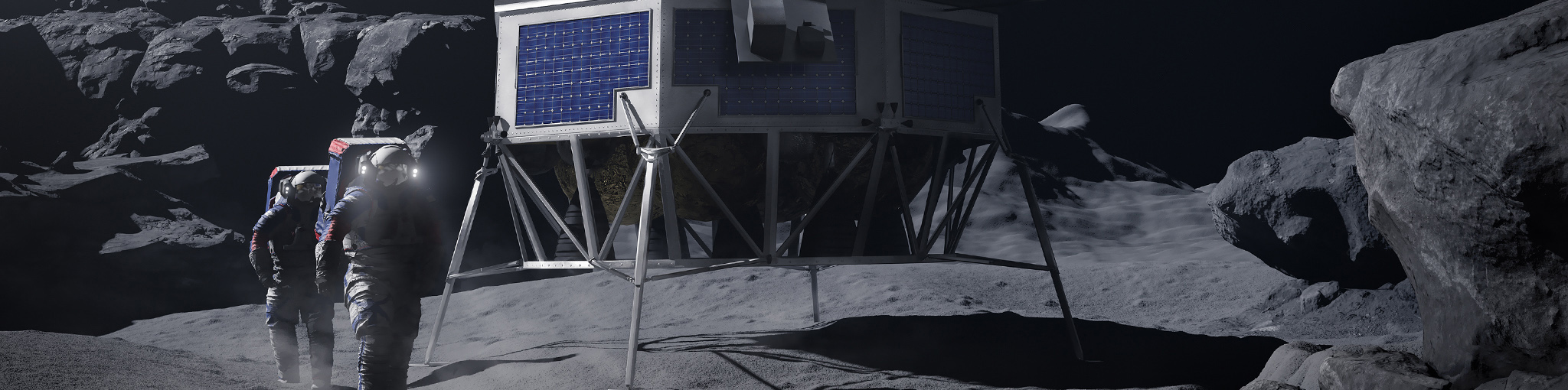}
  \caption{VR rendering of the lunar south pole, along with a prospective configuration of the EL3 lander.}
  \Description{VR recreation of the lunar south pole, along with a prospective configuration of the EL3 lander.}
  \label{fig:teaser}
\end{teaserfigure}

\maketitle

\section{Introduction}
50 years since humans last walked on the Moon, a renewed international interest coalescing around the Artemis programme seeks to send astronauts back to the lunar surface, aiming to establish a permanent human presence on the Moon by 2028 \cite{smith2020artemis}. The feasibility of this undertaking will be preconditioned by the development of reliable logistics supply solutions in support of human expeditions. To meet this need, the European Space Agency (ESA), along with partners from the industry and academia, are currently preparing the development of the European Large Logistics Lander (EL3). Specifically, the EL3 is being designed as a multipurpose autonomous lander capable of carrying out a range of prospective missions, including the delivery of crew supplies and heavy machinery, such as unpressurised rovers, to the lunar surface \cite{el3}. 

Designing such systems for human lunar missions demands taking into account the challenges and constraints posed by programmatic, physical and environmental factors. This means one must not only design for established terrestrial usability and ergonomics factors, but likewise account for a host of other conditions, including the lunar gravity, regolith (moondust) mitigation, limited field of view and range of motion due to extravehicular activity (EVA) suits, mental and physical fatigue, thermal and atmospheric conditions, limited capability for redundancy and extreme lighting conditions, such as pitch black shadows \cite{eppler1991lighting}.  

Lunar lighting conditions in particular have emerged as a prominent consideration during the ongoing design and preparatory activities. Notably, the tilt of the lunar axis creates a unique setting on the Moon's south pole, with elevated areas experiencing near permanent illumination, whilst the very low Sun angle simultaneously also blocks out all sunlight in recessed areas. Indeed, the access to persistent illumination (and the resulting availability of solar power), along with water ice preserved in permanently shadowed areas, was a key factor leading to the selection of the Moon’s south polar region as the likely location of the next human landing and the subsequent construction of the Artemis Base Camp \cite{artemis}.  

Given its ability to visualize any environment and interactively simulate hypothetical design solutions \cite{walch2017evaluating}, virtual reality (VR) stands, at least in theory, well positioned to help inform the design of EL3. At the same time, as noted by Aylward et al.,\textit{ VR research is still in its infancy and is lacking many critical components to evaluate the reliability, validity, and generalizability of its methods and results} \cite{aylward2021using}.  

Our work seeks to address this limitation. Herein we have produced a relevant VR-based scenario and evaluated it with experts in the field of human spaceflight. By reflecting on our preliminary findings, we provide an assessment of the potential offered by VR as an enabling technology in future studies of operational performance and human factors in the early stages of a lunar surface system’s design and development.

\section{Related Work}
Prior use of VR has demonstrated its viability as a research tool for studying user interactions with a range of prospective technologies, including  autonomous cars \cite{sportillo2017immersive}, healthcare solutions \cite{lohse2014virtual} and complex robotic systems \cite{miner1994interactive}.  
The benefits associated with the employment of VR span across fields and disciplines. For instance, VR-based engineering analyses and design reviews conducted early on in a development process have been found to enable designers to uncover potential product flaws which would have otherwise led to additional costs if manufactured \cite{jerald2015vr}. Similarly, VR-based interactive visualizations and simulations have proven useful in enabling the assessment of scenarios that would otherwise be too dangerous, expensive or impractical to evaluate in the real world \cite{hancock2008human}.  
Such enquiries have demonstrated the capacity of VR to elicit rich insights and help predict and assess a range of aspects surrounding prospective design solutions, including usability \cite{drey2020vrsketchin}, human factors and ergonomics \cite{wienrich2018assessing}, user's perceived sense of security \cite{somin2021breachmob} as well as acceptance levels \cite{fussell2021using}.  

Nevertheless, the extent to which findings made in VR may be applicable to the real world have frequently been contested. For example, studies comparing driving in VR simulators to that in the real world have found that perceived danger and immersion are lower in VR \cite{helland2013comparison}, while sleepiness appears to be higher \cite{hallvig2013sleepy}.  

Such findings can be seen as troubling in domains concerned with human safety. Engineers and designers in the space industry have to cope with particularly stringent reliability and safety requirements. Consequently, in spite of being prone to innovation, the space sector, and the human spaceflight domain especially, have been traditionally conservative in introducing novel design tools when compared to other industrial domains.

The first uses of VR in the context of human spaceflight were primarily serving training purposes. Notably, in 1993 NASA pioneered the use of this tool when training the Hubble space telescope flight team for a repair mission \cite{loftin1995training}.  

Not until the 21st century did VR and virtualization begin to gradually gain a firm foothold in space system design. Virtual environments have since then been employed, for instance, to help prevent misalignments and incorrect data flow in the International Space Station’s Columbus module \cite{cardano2009vr}, to assist with general satellite assembly troubleshooting \cite{geng2017virtual} and to analyze the replacement procedure of a Columbus cabin filter \cite{helin2017augmented}.  

A key strand of the Artemis programme advocates the creation of pathways to advancing low-technology readiness level (TRL) solutions and helping adapt these to the unique operational requirements of the coming moon landing endeavor \cite{artemis}. Following this call, VR has also been employed to validate ergonomics, assembly schemes and maintenance for the Orion lunar spacecraft \cite{via_satellite_2017}, as well as to collect astronaut feedback for an ergonomic assessment of the Gateway lunar space station modules \cite{thales_group_2021}. In a more direct application of user centered design at the earliest stages of planning and development, VR was likewise utilized to give users a chance to reflect on and influence the design of hypothetical future lunar outposts \cite{casini2018analysis}, as well as in-situ resource utilization solutions for the Moon \cite{maggiore2018fuel}. 

Our work seeks to carry on in this vein of inquiry. As laid out in the following section, by employing VR to evaluate the prospective EL3 lander with expert users, we aim to shed further light on its viability and validity as a design tool in this context.


\section{Methodology}

In order to assess an early digital concept design of the EL3 and to elicit actionable reflections that could realistically help steer its future design direction, our group adopted an approach centered around the use of VR scenarios \cite{carrol1999five, mackay2000video}. 

\subsection{The VR Scenario}
Our scenario consisted of two key building blocks: a virtual lunar landscape (or ‘moonscape’), and an agnostic 3D model of a theoretical EL3 configuration.  

Based on topographic maps captured by the Lunar Reconnaissance Orbiter \cite{smith2017summary}, we recreated virtually an area of 64 km\textsuperscript{2} in close vicinity of the Shackleton crater on the Lunar south pole (89.9°S 0.0°E). We selected this area due to having previously been identified as one of the candidate landing sites for the first Artemis human landing mission \cite{artemis}, thus providing a sufficiently plausible backdrop to our scenario. Due to the original map‘s pixel scale of 100 meters, smaller details (e.g. boulders) were added manually using 3D modeling software. The virtual moonscape was subsequently textured procedurally using a set of custom terrain shaders. This work was primarily guided by reference photos acquired from NASA's image gallery \cite{nasagallery}. Additionally, a lunar geologist from our organization assisted us during this process to ensure authenticity. Finally, the virtual environment was instantiated using the Unreal Engine 4 game engine. The sun was placed in the direction of north, at an angle of 1.5° above the horizon and its intensity set to 1.37 kW/m\textsuperscript{2} to mimic the lighting conditions on the lunar south pole \cite{vanoutryve2010analysis}. All forms of indirect lighting and light scattering were disabled in order to recreate the pitch black shadows stemming from the lack of lunar atmosphere.  

\begin{figure*}[ht]
\centering
\captionsetup{justification=centering}

\includegraphics[width=\textwidth]{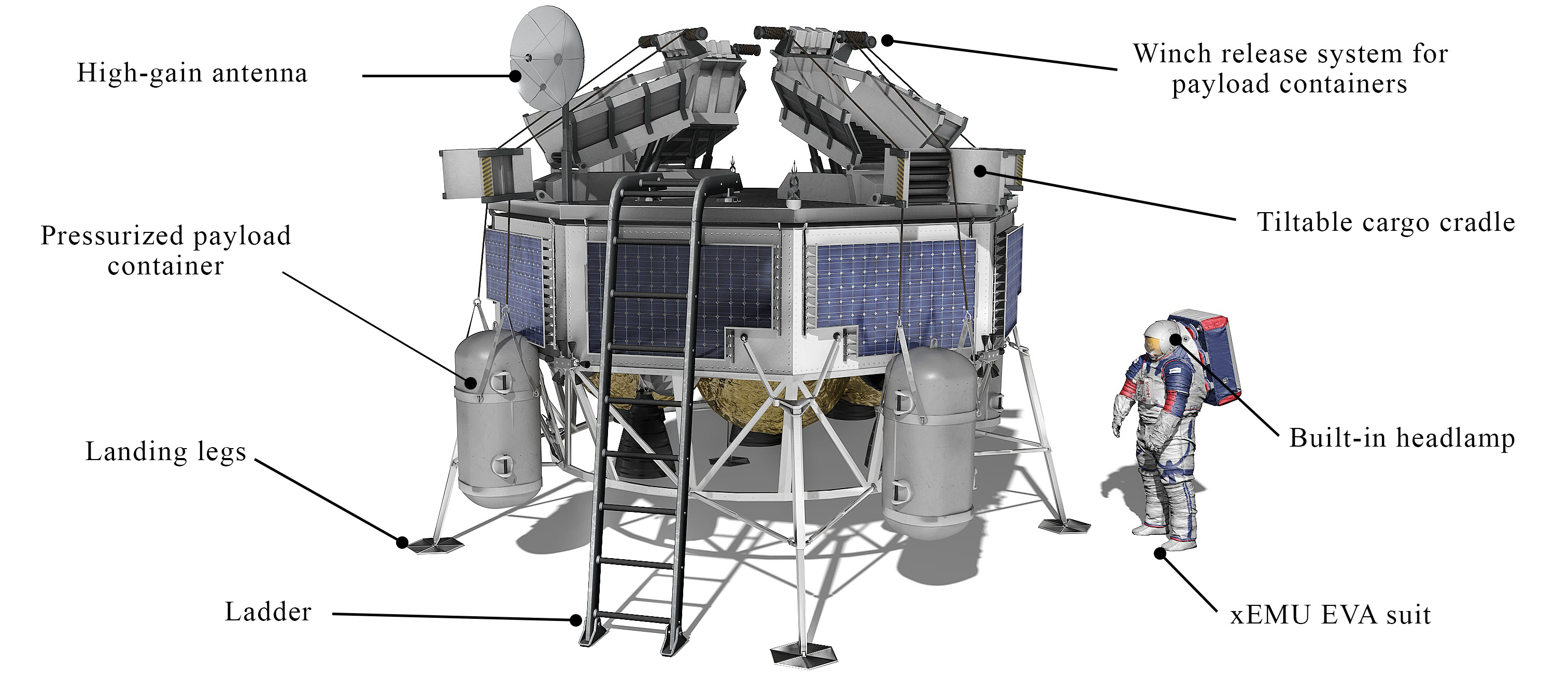}
\caption{3D model of the hypothetical EL3 configuration employed in our study }
\Description{3D model of the EL3 configuration employed in our study}
\label{figure:lander}
\vspace{5mm}
\end{figure*}

The development of our agnostic EL3 model was guided by a series of internal workshops in close collaboration with the EL3 project management and the responsible engineering teams. Based on their input, we created a model which represents a generic design that could meet the requirements outlined by the EL3’s invitation to tender \cite{Carey2021}. The design of the model was made to evaluate and explore general usability, operations and human factors challenges and constraints, which are likely to apply to most potential designs which might result from the EL3 team’s own work. On top of the virtual lander’s descent element, we designed and implemented a hypothetical cargo unloading solution in the form of a pulley system that would rely on lunar gravity to drop payload containers from the cargo deck at the top of the lander (see figure \ref{figure:lander}). The concept was validated via consultations with the EL3 team to assure our model could be used in a scenario which would provide actionable and relevant insights into the use of the lander and its cargo deployment system. Other potential solutions, including an autonomous robotic arm or a fully manual approach, were also discussed but not further pursued in this study. 

Our EL3 mockup carrying two payload containers was then placed in the middle of our virtual moonscape. A cargo drop-off point was placed 30 meters away from the lander and marked by a flag planted in the ground. Our scenario thus conveyed a basic, yet plausible, situation with users being tasked to approach the lander, retrieve the two containers and transport these to the nearby drop-off point.  

The VR experience was run on a desktop computer with an RTX 2080 GPU, providing smooth framerate. We used the Pimax VR headset coupled with a pair of HTC Vive base stations and controllers. Walking and turning was handled via the controller trackpads, while the controller triggers were used for interacting with objects in the virtual environment (e.g. lifting up payload containers). Users could also look around the environment by moving their head. To further improve authenticity of the scenario, users were embodied in a 3D model of a spacesuit. Specifically, we employed the Exploration Extravehicular Mobility Unit (xEMU) EVA suit \cite{ross2018nasa}. Users had their heads encapsulated inside the xEMU helmet to accurately restrict their field of view. The xEMU helmet was likewise equipped with a headlamp. This VR scenario then formed the focal point of our engagement with participants.

\subsection{Participants}
Given the niche character of our study, we recruited participants by hand-picking and extending invitations to staff and associates from ESA, ensuring that all participants had expertise in a field relevant to the subject of our inquiry. The study is currently ongoing, with 10 participants having completed a test session at the time of writing. These initial 10 participants are of varying levels of seniority, ranging from research students to International Space Station (ISS) ground segment engineers, and cover a spectrum of disciplines from space medicine to lunar science (see Table \ref{table:participants} for details). Only one of the 10 initial participants was female, future sessions will correct for this imbalance. None of the participants was directly involved in the design and development of the EL3. All of them did, however, have prior knowledge of the EL3 and were aware of the overarching goals it is being designed to meet.

\begin{center}
\begin{table*}[ht]
\small
\begin{tabular}{ |l|l|l|l| } 
\hline
\textbf{Participant} & \textbf{Gender} & \textbf{Job title} & \textbf{Expertise} \\
\hline
P1 & male & PhD student & Oxygen extraction from regolith\\ 

P2 & male & Intern & Regolith-based additive manufacturing \\ 
P3 & male & Science Advisor & Aerospace engineering \\ 
P4 & male & National trainee & Lunar in-situ resource utilization\\ 
P5 & male & Intern & Microwave regolith processing \\ 
P6 & male & National trainee & Space medicine research \\ 
P7 & male & ISS ground segment engineer & Spaceflight ops, avionics\\ 
P8 & male & Science and operations manager & Lunar science \\ 
P9 & male & ISS ground segment engineer & Aerospace engineering \\ 
P10 & female & Graduate trainee & Aerospace engineering \\ 
\hline
\end{tabular}
\caption{An overview of the study participants. Areas of expertise were self-reported.}
\Description{An overview of the study participants. The indicated areas of expertise were self-reported.}
    \label{table:participants}
    \vspace{-2mm}
    \end{table*}
\end{center}

\subsection{Procedure}
Participants were invited individually to complete our VR scenario. Every session began with the given participant being briefed about the purpose of our study and the task to be completed in the VR scenario, after which the participant was asked to provide consent to taking part in the study. This was followed by a quick demonstration of the VR controllers. Participants were then given freedom to navigate through the virtual environment and approach the container transportation objective in whichever way they preferred, as long as they did not stray too far away from the EL3 landing site. Drawing on the think aloud protocol \cite{ericsson1980verbal}, we encouraged our participants to verbalize their reasoning while completing the task. Once the payload containers were successfully retrieved and transported, we asked our participants to stay in the virtual environment for a little longer while answering a set of semi-structured interview questions. These questions were all open-ended and prompted participants to share their reflections and comment on any aspect of their experience. In particular we enquired about various features of the lunar lander, such as the cargo unloading mechanism, antenna placement, and the design of the ladder. Participants were also asked to identify potential safety hazards and suggest design improvements. Any unusual behaviors or actions exhibited by the participants were likewise explored. 

Finally, upon completing the semi-structured interview, participants were instructed to take off the VR headset and fill in a post-task questionnaire which quantitatively assessed their perceived mental workload using the NASA-TLX standard \cite{hart2006nasa} as well as their perceived presence in the virtual environment \cite{slater2018immersion}. This was to support a future comparative analysis. Due to the limited sample size at this stage of our study, quantitative data will however not be reported on in this paper.  

At the end of each session, the participant was debriefed and asked whether they would be happy to have their data included in the analysis. No time limits were applied during the study. Instead, we sought to provide each participant with as much (or as little) time as they needed to complete the scenario and to answer our questions. Consequently, the total length of the conducted sessions ranged widely from 20 to 60 minutes.  

\section{Preliminary Findings}
Participant responses were recorded in the form of notes, audio recordings and questionnaire replies. The dataset was independently coded by three of our researchers, with any inconsistencies being addressed through a discussion, and subsequently synthesized into a qualitative thematic analysis \cite{braun2006using}. 

The majority of our participants had either none or very limited experience with VR (self-reported on a 5-point scale). Nevertheless, they had no difficulty coming to grips with the VR controllers and completing the scenario. Only one participant (P10) reported discomfort in the form of dizziness caused by VR. All in all, the cargo transportation task was seen as very accessible, being frequently described as “straightforward” (P8) or “extremely easy” (P7). This, however, did not prevent our participants from making numerous reflections relevant to the EL3.  

\subsection{Key Challenges}

The unique lighting conditions on the lunar south pole were frequently brought up as the most immediate issue endangering the EL3 and the ground crew. The sun’s position near the lunar horizon resulted in a situation where every terrain feature casted protracted shadows, shrouding much of the environment in pitch darkness. As noted by P8, the shadows casted by players in VR were themselves deep enough to occlude any terrain inside them, which introduces the risk of stumbling on boulders or other surface objects. This became a particularly pressing problem when navigating directly away from the sun. Similarly, P3 argued that, should the lander be standing in a shadowed area, one could easily trip on its landing legs. P7 also noted that poor lighting conditions contribute to payload containers and other vital equipment being hard to find.  

While the built-in headlamp of the xEMU suit was seen as helpful, all participants agreed that in its current form it would be too weak to adequately support crews during extravehicular activity. P2 and P5, for instance, both favored making the headlamp stronger, while P10 advocated the addition of a secondary ‘chest lamp’ on the xEMU suit.  

Several features of the EL3 itself also appeared in need of being redesigned with greater visibility in mind. P6 argued that key elements of the lander should be highlighted via LED strips, reflective tapes or other warning signs. P3 proposed small low-energy light bars attached to the lander legs, while P9 argued that the EL3 should make better use of colors contrasting to the lunar environment, such as yellow or red, as a way of supporting astronaut orientation in poor lighting conditions. In order to make the EL3 more noticeable from greater distance, P4 and P10 also advocated it should be equipped with flashing navigation lights, not too different from those found on modern airplanes. 

The importance of equipping the EL3 with additional light sources got further amplified as our participants began to speculate about more complex interactions between astronauts and the EL3, such as the performance of various maintenance or repair tasks. Numerous hypothetical situations were postulated that would, for instance, require an astronaut to make use of the ladder and climb up on top of the EL3, including a conveyor belt jammed by regolith sediments (P1), a tangled cargo winch (P9) or a misaligned antenna (P10).  

Using the ladder to climb up on the cargo deck was seen as hazardous by all of our participants. Not just due to suited astronauts being at danger of tripping during their climb (P2) but also due to the risk of them unwittingly damaging equipment when operating in the cramped space on top of the lander (P5). To aggravate matters further, the near zero angle of sunlight resulted in half of the EL3 being occluded. When faced by the prospect of having to climb up on the ‘dark half’ of the lander, our participants agreed that appropriate placement of artificial lights would be critical. P8 and P9, for instance, advocated adding LED lights on the individual steps of the ladder, while P7 proposed two rings of lights looping around the lander to highlight its contour; one below the solar panels and another one above them.  

When examining the interplay of shadows and highlights on the EL3, as rendered in our VR scenario, P4 reached a somewhat different conclusion. Whilst he concurred that climbing up on the cargo deck from the shadowed side of the lander would indeed constitute a safety hazard, he argued that the source of this hazard would not be the absence of light (the headlamp would compensate for that adequately). Rather, the problem would be the sunlight which would hit astronauts in the face as soon as they reached the top of the ladder and gazed across the edge of the EL3 lander. In order to stop astronauts from getting blinded, P4 argued the helmet would need some kind of an automatic anti-glare visor that would be pulled down as a protection. The transition from pitch darkness to blinding light would in this sense be so fast, and come at such a sensitive moment, that future design efforts will likely need to take this into consideration.

\subsection{Criticism}
While our scenario did succeed in eliciting a range of suggestions concerning potential design improvements of the EL3, it is crucial to acknowledge that some of the technical limitations of our VR simulation likely also played a role in shaping the feedback we received. Indeed, our participants frequently pointed this out themselves.  

By way of example, existing VR headsets cannot fully recreate the blinding intensity of the sun as it would be experienced in a real lunar environment. There was, however, no clear consensus amongst our participants whether this limitation impedes the validity of our VR simulation. Whilst P4, for instance, argued that the simulated sun is bright enough to convey the idea, P8 argued it should be brighter still.  

The inability of our VR scenario to accurately simulate the full effects of low gravity was seen as an even greater problem. P3 and P4, for example, were adamant about the unrealistic ease of player movement. In reality, they explained, the reduced lunar gravity would force astronauts to hop around, rather than move smoothly. Walking would thus be significantly more challenging than what our simulation propounded.  

Similarly, the lack of haptic feedback emerged as another potentially significant limitation, as became evident when lifting and manipulating equipment, such as payload containers. Nearly all participants noted the lack of weight, with P1 going as far as stating that the containers feel as if "made from paper". Whereas our scenario made such interactions seem effortless, multiple participants noted that in reality they might require a major effort, or potentially even be prohibitively difficult to carry out. This was not simply due to the lack of simulated weight in VR, but also due to other factors, such as inertia, as P7 explained when reflecting on the notion of having to manually relocate the EL3 ladder: \textit{“Don't underestimate the fact that on the Moon weight is a lot less than on Earth, but the inertia is still there. You have the same problem in microgravity… have you seen the astronauts on ISS moving the big life support racks? It might seem easy, but in fact it’s hell. Because the rack is weightless, but it has an immense amount of inertia. Which means that if you need to stop its motion, or if you need to rotate it, it's difficult, it’s really difficult. And this is a big ladder. A big, big ladder. And I’m not sure whether you could comfortably take it out and put it somewhere else.” }

The lack of realism, at least in part ingrained in contemporary VR technology, was in this sense seen as risking to trivialize potentially significant design challenges surrounding lunar human-machine interactions, which could in worst case lead to their inadvertent derogation.    

\section{Discussion}
Virtual reality is enjoying growing popularity in a range of industries due to enabling fast, flexible and cost-efficient design processes. Following this trend, the ambition of our ongoing enquiry is to assess its viability in the unique context of the current lunar landing programme. Using the EL3 project as our arena, we have carried out a preliminary evaluation of a VR-based cargo unloading scenario featuring an expert group of participants.  

Our participant’s qualitative experiences have demonstrated the capacity of VR to elicit and invoke a range of potentially actionable reflections concerning ergonomics, safety and crew performance issues surrounding a low TRL concept. Whilst such findings are largely congruent with similarly aimed research that has taken place in other domains (e.g. \cite{goedicke2018vr}, \cite{reinschluessel2017virtual}), the lunar south pole setting introduces a number of unique factors. Predominantly, our study has surfaced a range of design considerations stemming from the lunar sui generis lighting conditions. Given that these lighting conditions will inevitably interface with nearly every facet of future human operations on the lunar surface, developing efficient means for their subsumption into forthcoming prototyping and design activities will be critical. With regards to our findings, then, we feel compelled to argue that the near absolute control over light behavior offered by modern VR engines, coupled with the opportunity to experience relevant scenarios immersively and interactively, uniquely qualifies VR for this particular purpose. 

Equally important, however, is to take into account some of the limitations impairing the use of VR in other areas. Accurate simulation of physical interactions in VR, for instance, remains a major technological challenge \cite{he2017physhare}. The lack of perceived lunar gravity, weight and inertia were explicitly identified as drawbacks by our participants and likely altered the outcomes of our study at least to some extent. Whether such deficits could partially invalidate the findings made during VR-based user studies is still up for debate. Nevertheless it does seem likely that they might leave a host of potential design challenges underexposed. There is a risk, in other words, that an overly strong reliance on VR may unintentionally brush over major human and technical problems, leading to misplaced or misguided design priorities.  
More work is therefore needed to better understand the impact of such deficits and to establish a clear rationale for employing virtual reality during the design of future lunar solutions. 

\subsection{Future Directions and the Problem of Validity }
Moving forward, our study will carry on with additional expert participants. This initial pilot testing phase will culminate late in Q1 2022, with our VR scenario being evaluated by astronauts with actual EVA experience. Such expert input will allow us to further elaborate and better understand potential utilization of VR in relevant user centered design activities.  

Moreover, our group is currently collaborating on the development of a 700 m\textsuperscript{2} lunar surface testbed area enabling high-fidelity replication of Moon surface conditions \cite{casini2020lunar}. The testbed facility will feature variable and adjustable lighting capable of simulating lunar illumination, along with a gravity offload system to mimic partial weight of astronauts or hardware. Relying on a life-size EL3 mockup, we aim to emulate our VR scenario inside the testbed environment and subsequently carry out a comparative analysis, with our original VR study serving as baseline.  

Such a comparative analysis will allow us to gain a better understanding of the strengths and weaknesses of virtual reality in comparison to more conventional design platforms. By exploring room for any potential synergies, we likewise hope to identify means to compensate for some of the limitations of VR, including the lack of physical interactions, by combining the two \cite{costantini2021extended, becker2021evaluation}. Indeed, studies such as Shaw et al. have shown that combining audiovisual stimulation with that of additional sensory modalities can measurably improve validity of observations made in VR scenarios \cite{shaw2019heat}. The notion of fusing real and virtual environments has already shown promise in practice. Military pilots, for instance, have made use of VR headsets while flying real planes to train tasks such as mid-flight refueling \cite{conner2015fused}. Ideally, such forms of mixed reality would in this sense merge the best of both worlds, facilitating more authentic experiences than what could be achieved using VR alone. 

\section{Conclusions}
Our evaluation of a prototypical implementation of the EL3 has surfaced unique capabilities possessed by VR in the context of lunar surface system design. In particular, we have argued that the capacity of VR to accurately simulate lighting conditions is alone a sufficiently powerful incentive for its employment in relevant design activities. On the other hand, we likewise came across situations where technical limitations of VR appeared to impede the validity of participant's observations. We have argued that additional modes of sensory stimulation should be explored to compensate for this. Although it may never be feasible to fully recreate the extreme conditions of the lunar south pole in a lab, by identifying and efficiently amalgamating the strengths of available tools and methods, we may provide designers with the best means possible to elicit valid reflections and consequently better anticipate the many contingencies faced by future lunar expeditions. It is the aspiration of our work to bring us closer to this goal.

\begin{acks}
We would like to thank the European Astronaut Center in Cologne for providing us with expert participants for our study. We are also thankful to Ludovic Duvet and his EL3 team for overseeing our prototyping activities.  
\end{acks}

\balance

\bibliographystyle{ACM-Reference-Format}
\bibliography{main.bib}


\begin{thebibliography}{44}


\ifx \showCODEN    \undefined \def \showCODEN     #1{\unskip}     \fi
\ifx \showDOI      \undefined \def \showDOI       #1{#1}\fi
\ifx \showISBNx    \undefined \def \showISBNx     #1{\unskip}     \fi
\ifx \showISBNxiii \undefined \def \showISBNxiii  #1{\unskip}     \fi
\ifx \showISSN     \undefined \def \showISSN      #1{\unskip}     \fi
\ifx \showLCCN     \undefined \def \showLCCN      #1{\unskip}     \fi
\ifx \shownote     \undefined \def \shownote      #1{#1}          \fi
\ifx \showarticletitle \undefined \def \showarticletitle #1{#1}   \fi
\ifx \showURL      \undefined \def \showURL       {\relax}        \fi
\providecommand\bibfield[2]{#2}
\providecommand\bibinfo[2]{#2}
\providecommand\natexlab[1]{#1}
\providecommand\showeprint[2][]{arXiv:#2}

\bibitem[Aylward et~al\mbox{.}(2021)]%
        {aylward2021using}
\bibfield{author}{\bibinfo{person}{Katie Aylward}, \bibinfo{person}{Joakim
  Dahlman}, \bibinfo{person}{Kjetil Nordby}, {and} \bibinfo{person}{Monica
  Lundh}.} \bibinfo{year}{2021}\natexlab{}.
\newblock \showarticletitle{Using operational scenarios in a virtual reality
  enhanced design process}.
\newblock \bibinfo{journal}{\emph{Education Sciences}} \bibinfo{volume}{11},
  \bibinfo{number}{8} (\bibinfo{year}{2021}), \bibinfo{pages}{448}.
\newblock


\bibitem[Becker(2021)]%
        {becker2021evaluation}
\bibfield{author}{\bibinfo{person}{Leonie Becker}.}
  \bibinfo{year}{2021}\natexlab{}.
\newblock \emph{\bibinfo{title}{Evaluation of Haptic Collision Feedback and
  Haptic Guidance for Robotic Teleoperation}}.
\newblock \bibinfo{thesistype}{Ph.\,D. Dissertation}.
  \bibinfo{school}{Universit{\"a}t Siegen}.
\newblock


\bibitem[Braun and Clarke(2006)]%
        {braun2006using}
\bibfield{author}{\bibinfo{person}{Virginia Braun} {and}
  \bibinfo{person}{Victoria Clarke}.} \bibinfo{year}{2006}\natexlab{}.
\newblock \showarticletitle{Using thematic analysis in psychology}.
\newblock \bibinfo{journal}{\emph{Qualitative research in psychology}}
  \bibinfo{volume}{3}, \bibinfo{number}{2} (\bibinfo{year}{2006}),
  \bibinfo{pages}{77--101}.
\newblock


\bibitem[Cardano et~al\mbox{.}(2009)]%
        {cardano2009vr}
\bibfield{author}{\bibinfo{person}{M Cardano}, \bibinfo{person}{M Ferrino},
  \bibinfo{person}{M Costa}, {and} \bibinfo{person}{P Giorgi}.}
  \bibinfo{year}{2009}\natexlab{}.
\newblock \showarticletitle{VR/AR tools to support on orbit crew operations and
  P/Ls maintenance in the ISS pressurized Columbus module}. In
  \bibinfo{booktitle}{\emph{60th International Astronautical Congress, Daejeon,
  Republic of Korea}}. \bibinfo{pages}{12--16}.
\newblock


\bibitem[Carey et~al\mbox{.}(2021)]%
        {Carey2021}
\bibfield{author}{\bibinfo{person}{William~C. Carey}, \bibinfo{person}{Ludovic
  Duvet}, \bibinfo{person}{Nick Gollins}, \bibinfo{person}{Rogier Schonenborg},
  \bibinfo{person}{Alexander Cropp}, \bibinfo{person}{Giorgio Cifani},
  \bibinfo{person}{Keith Stephenson}, \bibinfo{person}{Philipp~B. Hager},
  \bibinfo{person}{Yannick~Le Deuff}, \bibinfo{person}{Kim Nergaard},
  \bibinfo{person}{Thorsten Graber}, \bibinfo{person}{Jennifer Reynolds},
  \bibinfo{person}{Giorgio Magistrati}, \bibinfo{person}{Sandra Magunsong},
  \bibinfo{person}{Jorge Alves}, \bibinfo{person}{Francesca McDonald}, {and}
  \bibinfo{person}{Nadine Boersma}.} \bibinfo{year}{2021}\natexlab{}.
\newblock \showarticletitle{European access to the lunar surface: EL3 mission
  options}.
\newblock \bibinfo{journal}{\emph{72nd International Astronautical Congress
  (IAC)}}.
\newblock


\bibitem[Carrol(1999)]%
        {carrol1999five}
\bibfield{author}{\bibinfo{person}{John~M Carrol}.}
  \bibinfo{year}{1999}\natexlab{}.
\newblock \showarticletitle{Five reasons for scenario-based design}. In
  \bibinfo{booktitle}{\emph{Proceedings of the 32nd Annual Hawaii International
  Conference on Systems Sciences. 1999. HICSS-32. Abstracts and CD-ROM of Full
  Papers}}. IEEE, \bibinfo{pages}{11--pp}.
\newblock


\bibitem[Casini et~al\mbox{.}(2018)]%
        {casini2018analysis}
\bibfield{author}{\bibinfo{person}{Andrea~EM Casini}, \bibinfo{person}{Paolo
  Maggiore}, \bibinfo{person}{Nicole Viola}, \bibinfo{person}{Valter Basso},
  \bibinfo{person}{Marinella Ferrino}, \bibinfo{person}{Jeffrey~A Hoffman},
  {and} \bibinfo{person}{Aidan Cowley}.} \bibinfo{year}{2018}\natexlab{}.
\newblock \showarticletitle{Analysis of a Moon outpost for Mars enabling
  technologies through a Virtual Reality environment}.
\newblock \bibinfo{journal}{\emph{Acta Astronautica}}  \bibinfo{volume}{143}
  (\bibinfo{year}{2018}), \bibinfo{pages}{353--361}.
\newblock


\bibitem[Casini et~al\mbox{.}(2020)]%
        {casini2020lunar}
\bibfield{author}{\bibinfo{person}{Andrea~EM Casini}, \bibinfo{person}{Petra
  Mittler}, \bibinfo{person}{Aidan Cowley}, \bibinfo{person}{Lukas
  Schl{\"u}ter}, \bibinfo{person}{Marthe Faber}, \bibinfo{person}{Beate
  Fischer}, \bibinfo{person}{Melanie von~der Wiesche}, {and}
  \bibinfo{person}{Matthias Maurer}.} \bibinfo{year}{2020}\natexlab{}.
\newblock \showarticletitle{Lunar analogue facilities development at EAC: the
  LUNA project}.
\newblock \bibinfo{journal}{\emph{Journal of Space Safety Engineering}}
  \bibinfo{volume}{7}, \bibinfo{number}{4} (\bibinfo{year}{2020}),
  \bibinfo{pages}{510--518}.
\newblock


\bibitem[Conner(2015)]%
        {conner2015fused}
\bibfield{author}{\bibinfo{person}{Monroe Conner}.}
  \bibinfo{year}{2015}\natexlab{}.
\newblock \bibinfo{title}{Fused Reality: Making the Imagined Seem Real}.
\newblock
\newblock


\bibitem[Costantini et~al\mbox{.}(2021)]%
        {costantini2021extended}
\bibfield{author}{\bibinfo{person}{Martial Costantini}, \bibinfo{person}{Flavie
  Rometsch}, \bibinfo{person}{Andrea Emanuele~Maria Casini},
  \bibinfo{person}{Aidan Cowley}, \bibinfo{person}{Stephen Ennis},
  \bibinfo{person}{Christopher Scott}, \bibinfo{person}{Stephan Ghiste},
  \bibinfo{person}{Jonathan Scott}, {and} \bibinfo{person}{Lionel Ferra}.}
  \bibinfo{year}{2021}\natexlab{}.
\newblock \showarticletitle{eXtended Reality applications for human
  spaceflight: the ESA-EAC XR Lab}. In \bibinfo{booktitle}{\emph{Proceedings of
  the International Astronautical Congress, IAC}}.
\newblock


\bibitem[Drey et~al\mbox{.}(2020)]%
        {drey2020vrsketchin}
\bibfield{author}{\bibinfo{person}{Tobias Drey}, \bibinfo{person}{Jan
  Gugenheimer}, \bibinfo{person}{Julian Karlbauer}, \bibinfo{person}{Maximilian
  Milo}, {and} \bibinfo{person}{Enrico Rukzio}.}
  \bibinfo{year}{2020}\natexlab{}.
\newblock \showarticletitle{Vrsketchin: Exploring the design space of pen and
  tablet interaction for 3d sketching in virtual reality}. In
  \bibinfo{booktitle}{\emph{Proceedings of the 2020 CHI Conference on Human
  Factors in Computing Systems}}. \bibinfo{pages}{1--14}.
\newblock


\bibitem[Eppler(1991)]%
        {eppler1991lighting}
\bibfield{author}{\bibinfo{person}{Dean~B Eppler}.}
  \bibinfo{year}{1991}\natexlab{}.
\newblock \showarticletitle{Lighting constraints on lunar surface operations}.
\newblock \bibinfo{journal}{\emph{NASA STI/Recon Technical Report N}}
  \bibinfo{volume}{91} (\bibinfo{year}{1991}), \bibinfo{pages}{23014}.
\newblock


\bibitem[Ericsson and Simon(1980)]%
        {ericsson1980verbal}
\bibfield{author}{\bibinfo{person}{K~Anders Ericsson} {and}
  \bibinfo{person}{Herbert~A Simon}.} \bibinfo{year}{1980}\natexlab{}.
\newblock \showarticletitle{Verbal reports as data.}
\newblock \bibinfo{journal}{\emph{Psychological review}} \bibinfo{volume}{87},
  \bibinfo{number}{3} (\bibinfo{year}{1980}), \bibinfo{pages}{215}.
\newblock


\bibitem[Fussell and Truong(2021)]%
        {fussell2021using}
\bibfield{author}{\bibinfo{person}{Stephanie~G Fussell} {and}
  \bibinfo{person}{Dothang Truong}.} \bibinfo{year}{2021}\natexlab{}.
\newblock \showarticletitle{Using virtual reality for dynamic learning: an
  extended technology acceptance model}.
\newblock \bibinfo{journal}{\emph{Virtual Reality}} (\bibinfo{year}{2021}),
  \bibinfo{pages}{1--19}.
\newblock


\bibitem[Geng et~al\mbox{.}(2017)]%
        {geng2017virtual}
\bibfield{author}{\bibinfo{person}{Jie Geng}, \bibinfo{person}{Ying Li},
  \bibinfo{person}{Ranran Wang}, \bibinfo{person}{Zili Wang},
  \bibinfo{person}{Chuan Lv}, {and} \bibinfo{person}{Dong Zhou}.}
  \bibinfo{year}{2017}\natexlab{}.
\newblock \showarticletitle{A virtual maintenance-based approach for satellite
  assembling and troubleshooting assessment}.
\newblock \bibinfo{journal}{\emph{Acta Astronautica}}  \bibinfo{volume}{138}
  (\bibinfo{year}{2017}), \bibinfo{pages}{434--453}.
\newblock


\bibitem[Goedicke et~al\mbox{.}(2018)]%
        {goedicke2018vr}
\bibfield{author}{\bibinfo{person}{David Goedicke}, \bibinfo{person}{Jamy Li},
  \bibinfo{person}{Vanessa Evers}, {and} \bibinfo{person}{Wendy Ju}.}
  \bibinfo{year}{2018}\natexlab{}.
\newblock \showarticletitle{Vr-oom: Virtual reality on-road driving
  simulation}. In \bibinfo{booktitle}{\emph{Proceedings of the 2018 CHI
  Conference on Human Factors in Computing Systems}}. \bibinfo{pages}{1--11}.
\newblock


\bibitem[Gollins et~al\mbox{.}(2020)]%
        {el3}
\bibfield{author}{\bibinfo{person}{Nick Gollins}, \bibinfo{person}{Shahrzad
  Timman}, \bibinfo{person}{Max Braun}, {and} \bibinfo{person}{Markus
  Landgraf}.} \bibinfo{year}{2020}\natexlab{}.
\newblock \showarticletitle{Building a European Lunar Capability with the
  European Large Logistic Lander}. In \bibinfo{booktitle}{\emph{EGU General
  Assembly Conference Abstracts}}. \bibinfo{pages}{22568}.
\newblock


\bibitem[Hallvig et~al\mbox{.}(2013)]%
        {hallvig2013sleepy}
\bibfield{author}{\bibinfo{person}{David Hallvig}, \bibinfo{person}{Anna
  Anund}, \bibinfo{person}{Carina Fors}, \bibinfo{person}{G{\"o}ran Kecklund},
  \bibinfo{person}{Johan~G Karlsson}, \bibinfo{person}{Mattias Wahde}, {and}
  \bibinfo{person}{Torbj{\"o}rn {\AA}kerstedt}.}
  \bibinfo{year}{2013}\natexlab{}.
\newblock \showarticletitle{Sleepy driving on the real road and in the
  simulator—A comparison}.
\newblock \bibinfo{journal}{\emph{Accident Analysis \& Prevention}}
  \bibinfo{volume}{50} (\bibinfo{year}{2013}), \bibinfo{pages}{44--50}.
\newblock


\bibitem[Hancock et~al\mbox{.}(2008)]%
        {hancock2008human}
\bibfield{author}{\bibinfo{person}{Peter~A Hancock}, \bibinfo{person}{Dennis~A
  Vincenzi}, \bibinfo{person}{John~A Wise}, {and} \bibinfo{person}{Mustapha
  Mouloua}.} \bibinfo{year}{2008}\natexlab{}.
\newblock \bibinfo{booktitle}{\emph{Human factors in simulation and training}}.
\newblock \bibinfo{publisher}{CRC Press}.
\newblock


\bibitem[Hart(2006)]%
        {hart2006nasa}
\bibfield{author}{\bibinfo{person}{Sandra~G Hart}.}
  \bibinfo{year}{2006}\natexlab{}.
\newblock \showarticletitle{NASA-task load index (NASA-TLX); 20 years later}.
  In \bibinfo{booktitle}{\emph{Proceedings of the human factors and ergonomics
  society annual meeting}}, Vol.~\bibinfo{volume}{50}. Sage publications Sage
  CA: Los Angeles, CA, \bibinfo{pages}{904--908}.
\newblock


\bibitem[He et~al\mbox{.}(2017)]%
        {he2017physhare}
\bibfield{author}{\bibinfo{person}{Zhenyi He}, \bibinfo{person}{Fengyuan Zhu},
  {and} \bibinfo{person}{Ken Perlin}.} \bibinfo{year}{2017}\natexlab{}.
\newblock \showarticletitle{Physhare: Sharing physical interaction in virtual
  reality}. In \bibinfo{booktitle}{\emph{Adjunct Publication of the 30th Annual
  ACM Symposium on User Interface Software and Technology}}.
  \bibinfo{pages}{17--19}.
\newblock


\bibitem[Helin(2017)]%
        {helin2017augmented}
\bibfield{author}{\bibinfo{person}{Kaj Helin}.}
  \bibinfo{year}{2017}\natexlab{}.
\newblock \showarticletitle{Augmented reality for AIT, AIV and operations}. In
  \bibinfo{booktitle}{\emph{Space Engineering and Technology Final Presentation
  Days}}.
\newblock


\bibitem[Helland et~al\mbox{.}(2013)]%
        {helland2013comparison}
\bibfield{author}{\bibinfo{person}{Arne Helland}, \bibinfo{person}{Gunnar~D
  Jenssen}, \bibinfo{person}{Lone-Eirin Lerv{\aa}g},
  \bibinfo{person}{Andreas~Austgulen Westin}, \bibinfo{person}{Terje Moen},
  \bibinfo{person}{Kristian Sakshaug}, \bibinfo{person}{Stian Lydersen},
  \bibinfo{person}{J{\o}rg M{\o}rland}, {and} \bibinfo{person}{Lars
  Sl{\o}rdal}.} \bibinfo{year}{2013}\natexlab{}.
\newblock \showarticletitle{Comparison of driving simulator performance with
  real driving after alcohol intake: A randomised, single blind,
  placebo-controlled, cross-over trial}.
\newblock \bibinfo{journal}{\emph{Accident Analysis \& Prevention}}
  \bibinfo{volume}{53} (\bibinfo{year}{2013}), \bibinfo{pages}{9--16}.
\newblock


\bibitem[Jerald(2015)]%
        {jerald2015vr}
\bibfield{author}{\bibinfo{person}{Jason Jerald}.}
  \bibinfo{year}{2015}\natexlab{}.
\newblock \bibinfo{booktitle}{\emph{The VR book: Human-centered design for
  virtual reality}}.
\newblock \bibinfo{publisher}{Morgan \& Claypool}.
\newblock


\bibitem[Loftin and Kenney(1995)]%
        {loftin1995training}
\bibfield{author}{\bibinfo{person}{R~Bowen Loftin} {and} \bibinfo{person}{P
  Kenney}.} \bibinfo{year}{1995}\natexlab{}.
\newblock \showarticletitle{Training the Hubble space telescope flight team}.
\newblock \bibinfo{journal}{\emph{IEEE Computer Graphics and Applications}}
  \bibinfo{volume}{15}, \bibinfo{number}{5} (\bibinfo{year}{1995}),
  \bibinfo{pages}{31--37}.
\newblock


\bibitem[Lohse et~al\mbox{.}(2014)]%
        {lohse2014virtual}
\bibfield{author}{\bibinfo{person}{Keith~R Lohse}, \bibinfo{person}{Courtney~GE
  Hilderman}, \bibinfo{person}{Katharine~L Cheung}, \bibinfo{person}{Sandy
  Tatla}, {and} \bibinfo{person}{HF~Machiel Van~der Loos}.}
  \bibinfo{year}{2014}\natexlab{}.
\newblock \showarticletitle{Virtual reality therapy for adults post-stroke: a
  systematic review and meta-analysis exploring virtual environments and
  commercial games in therapy}.
\newblock \bibinfo{journal}{\emph{PloS one}} \bibinfo{volume}{9},
  \bibinfo{number}{3} (\bibinfo{year}{2014}), \bibinfo{pages}{e93318}.
\newblock


\bibitem[Mackay et~al\mbox{.}(2000)]%
        {mackay2000video}
\bibfield{author}{\bibinfo{person}{Wendy~E Mackay}, \bibinfo{person}{Anne~V
  Ratzer}, {and} \bibinfo{person}{Paul Janecek}.}
  \bibinfo{year}{2000}\natexlab{}.
\newblock \showarticletitle{Video artifacts for design: Bridging the gap
  between abstraction and detail}. In \bibinfo{booktitle}{\emph{Proceedings of
  the 3rd conference on Designing interactive systems: processes, practices,
  methods, and techniques}}. \bibinfo{pages}{72--82}.
\newblock


\bibitem[Maggiore et~al\mbox{.}(2018)]%
        {maggiore2018fuel}
\bibfield{author}{\bibinfo{person}{Paolo Maggiore}, \bibinfo{person}{Alessio
  Cataudella}, {and} \bibinfo{person}{Andrea Emanuele~Maria Casini}.}
  \bibinfo{year}{2018}\natexlab{}.
\newblock \showarticletitle{Fuel cell test rig reconfiguration for a space
  energy provision system}.
\newblock  (\bibinfo{year}{2018}).
\newblock


\bibitem[Miner and Stansfield(1994)]%
        {miner1994interactive}
\bibfield{author}{\bibinfo{person}{Nadine~E Miner} {and}
  \bibinfo{person}{Sharon~A Stansfield}.} \bibinfo{year}{1994}\natexlab{}.
\newblock \showarticletitle{An interactive virtual reality simulation system
  for robot control and operator training}. In
  \bibinfo{booktitle}{\emph{Proceedings of the 1994 IEEE International
  Conference on Robotics and Automation}}. IEEE, \bibinfo{pages}{1428--1435}.
\newblock


\bibitem[NASA(2022)]%
        {nasagallery}
\bibfield{author}{\bibinfo{person}{NASA}.} \bibinfo{year}{2022}\natexlab{}.
\newblock \bibinfo{booktitle}{\emph{{NASA Image Gallery}}}.
\newblock
\urldef\tempurl%
\url{https://images.nasa.gov/}
\showURL{%
\tempurl}


\bibitem[Reinschluessel et~al\mbox{.}(2017)]%
        {reinschluessel2017virtual}
\bibfield{author}{\bibinfo{person}{Anke~Verena Reinschluessel},
  \bibinfo{person}{Joern Teuber}, \bibinfo{person}{Marc Herrlich},
  \bibinfo{person}{Jeffrey Bissel}, \bibinfo{person}{Melanie van Eikeren},
  \bibinfo{person}{Johannes Ganser}, \bibinfo{person}{Felicia Koeller},
  \bibinfo{person}{Fenja Kollasch}, \bibinfo{person}{Thomas Mildner},
  \bibinfo{person}{Luca Raimondo}, {et~al\mbox{.}}}
  \bibinfo{year}{2017}\natexlab{}.
\newblock \showarticletitle{Virtual reality for user-centered design and
  evaluation of touch-free interaction techniques for navigating medical images
  in the operating room}. In \bibinfo{booktitle}{\emph{Proceedings of the 2017
  CHI Conference Extended Abstracts on Human Factors in Computing Systems}}.
  \bibinfo{pages}{2001--2009}.
\newblock


\bibitem[Ross et~al\mbox{.}(2018)]%
        {ross2018nasa}
\bibfield{author}{\bibinfo{person}{Amy Ross}, \bibinfo{person}{Richard Rhodes},
  {and} \bibinfo{person}{Shane McFarland}.} \bibinfo{year}{2018}\natexlab{}.
\newblock \showarticletitle{NASA’s advanced extra-vehicular activity space
  suit pressure garment 2018 status and development plan}. 48th International
  Conference on Environmental Systems.
\newblock


\bibitem[Russell(2017)]%
        {via_satellite_2017}
\bibfield{author}{\bibinfo{person}{Kendall Russell}.}
  \bibinfo{year}{2017}\natexlab{}.
\newblock \bibinfo{booktitle}{\emph{{Lockheed Martin on cutting costs with
  virtual reality - via satellite}}}.
\newblock
\urldef\tempurl%
\url{https://www.satellitetoday.com/innovation/2017/04/20/lockheed-martin-cutting-costs-virtual-reality/}
\showURL{%
\tempurl}


\bibitem[Shaw et~al\mbox{.}(2019)]%
        {shaw2019heat}
\bibfield{author}{\bibinfo{person}{Emily Shaw}, \bibinfo{person}{Tessa Roper},
  \bibinfo{person}{Tommy Nilsson}, \bibinfo{person}{Glyn Lawson},
  \bibinfo{person}{Sue~VG Cobb}, {and} \bibinfo{person}{Daniel Miller}.}
  \bibinfo{year}{2019}\natexlab{}.
\newblock \showarticletitle{The heat is on: Exploring user behaviour in a
  multisensory virtual environment for fire evacuation}. In
  \bibinfo{booktitle}{\emph{Proceedings of the 2019 CHI Conference on Human
  Factors in Computing Systems}}. \bibinfo{pages}{1--13}.
\newblock


\bibitem[Slater(2018)]%
        {slater2018immersion}
\bibfield{author}{\bibinfo{person}{Mel Slater}.}
  \bibinfo{year}{2018}\natexlab{}.
\newblock \showarticletitle{Immersion and the illusion of presence in virtual
  reality}.
\newblock \bibinfo{journal}{\emph{British Journal of Psychology}}
  \bibinfo{volume}{109}, \bibinfo{number}{3} (\bibinfo{year}{2018}),
  \bibinfo{pages}{431--433}.
\newblock


\bibitem[Smith et~al\mbox{.}(2017)]%
        {smith2017summary}
\bibfield{author}{\bibinfo{person}{David~E Smith}, \bibinfo{person}{Maria~T
  Zuber}, \bibinfo{person}{Gregory~A Neumann}, \bibinfo{person}{Erwan
  Mazarico}, \bibinfo{person}{Frank~G Lemoine}, \bibinfo{person}{James~W
  Head~III}, \bibinfo{person}{Paul~G Lucey}, \bibinfo{person}{Oded Aharonson},
  \bibinfo{person}{Mark~S Robinson}, \bibinfo{person}{Xiaoli Sun},
  {et~al\mbox{.}}} \bibinfo{year}{2017}\natexlab{}.
\newblock \showarticletitle{Summary of the results from the lunar orbiter laser
  altimeter after seven years in lunar orbit}.
\newblock \bibinfo{journal}{\emph{Icarus}}  \bibinfo{volume}{283}
  (\bibinfo{year}{2017}), \bibinfo{pages}{70--91}.
\newblock


\bibitem[Smith et~al\mbox{.}(2020)]%
        {smith2020artemis}
\bibfield{author}{\bibinfo{person}{Marshall Smith}, \bibinfo{person}{Douglas
  Craig}, \bibinfo{person}{Nicole Herrmann}, \bibinfo{person}{Erin Mahoney},
  \bibinfo{person}{Jonathan Krezel}, \bibinfo{person}{Nate McIntyre}, {and}
  \bibinfo{person}{Kandyce Goodliff}.} \bibinfo{year}{2020}\natexlab{}.
\newblock \showarticletitle{The artemis program: An overview of nasa's
  activities to return humans to the moon}. In \bibinfo{booktitle}{\emph{2020
  IEEE Aerospace Conference}}. IEEE, \bibinfo{pages}{1--10}.
\newblock


\bibitem[Somin et~al\mbox{.}(2021)]%
        {somin2021breachmob}
\bibfield{author}{\bibinfo{person}{Lior Somin}, \bibinfo{person}{Zachary
  McKendrick}, \bibinfo{person}{Patrick Finn}, {and} \bibinfo{person}{Ehud
  Sharlin}.} \bibinfo{year}{2021}\natexlab{}.
\newblock \showarticletitle{BreachMob: Detecting Vulnerabilities in Physical
  Environments Using Virtual Reality}. In \bibinfo{booktitle}{\emph{Proceedings
  of the 27th ACM Symposium on Virtual Reality Software and Technology}}.
  \bibinfo{pages}{1--6}.
\newblock


\bibitem[Space(2021)]%
        {thales_group_2021}
\bibfield{author}{\bibinfo{person}{Thales~Alenia Space}.}
  \bibinfo{year}{2021}\natexlab{}.
\newblock \bibinfo{booktitle}{\emph{{ESA astronauts Alexander Gerst and Luca
  Parmitano use their avatars to check out future habitation accommodation on
  Lunar Gateway}}}.
\newblock
\urldef\tempurl%
\url{https://www.thalesgroup.com/en/worldwide/space/news/esa-astronauts-alexander-gerst-and-luca-parmitano-use-their-avatars-check-out}
\showURL{%
\tempurl}


\bibitem[Sportillo et~al\mbox{.}(2017)]%
        {sportillo2017immersive}
\bibfield{author}{\bibinfo{person}{Daniele Sportillo}, \bibinfo{person}{Alexis
  Paljic}, \bibinfo{person}{Mehdi Boukhris}, \bibinfo{person}{Philippe Fuchs},
  \bibinfo{person}{Luciano Ojeda}, {and} \bibinfo{person}{Vincent Roussarie}.}
  \bibinfo{year}{2017}\natexlab{}.
\newblock \showarticletitle{An immersive Virtual Reality system for
  semi-autonomous driving simulation: a comparison between realistic and 6-DoF
  controller-based interaction}. In \bibinfo{booktitle}{\emph{Proceedings of
  the 9th International Conference on Computer and Automation Engineering}}.
  \bibinfo{pages}{6--10}.
\newblock


\bibitem[Vanoutryve et~al\mbox{.}(2010)]%
        {vanoutryve2010analysis}
\bibfield{author}{\bibinfo{person}{Benjamin Vanoutryve}, \bibinfo{person}{Diego
  De~Rosa}, \bibinfo{person}{Richard Fisackerly}, \bibinfo{person}{Berengere
  Houdou}, \bibinfo{person}{James Carpenter}, \bibinfo{person}{Christian
  Philippe}, \bibinfo{person}{Alain Pradier}, \bibinfo{person}{Aliac
  Jojaghaian}, \bibinfo{person}{Sylvie Espinasse}, {and} \bibinfo{person}{Bruno
  Gardini}.} \bibinfo{year}{2010}\natexlab{}.
\newblock \showarticletitle{An analysis of illumination and communication
  conditions near lunar south pole based on Kaguya Data}. In
  \bibinfo{booktitle}{\emph{Proceedings of International Planetary Probe
  Workshop, Barcelona}}.
\newblock


\bibitem[Walch et~al\mbox{.}(2017)]%
        {walch2017evaluating}
\bibfield{author}{\bibinfo{person}{Marcel Walch}, \bibinfo{person}{Julian
  Frommel}, \bibinfo{person}{Katja Rogers}, \bibinfo{person}{Felix
  Sch{\"u}ssel}, \bibinfo{person}{Philipp Hock}, \bibinfo{person}{David
  Dobbelstein}, {and} \bibinfo{person}{Michael Weber}.}
  \bibinfo{year}{2017}\natexlab{}.
\newblock \showarticletitle{Evaluating VR driving simulation from a player
  experience perspective}. In \bibinfo{booktitle}{\emph{Proceedings of the 2017
  CHI Conference Extended Abstracts on Human Factors in Computing Systems}}.
  \bibinfo{pages}{2982--2989}.
\newblock


\bibitem[Weber et~al\mbox{.}(2021)]%
        {artemis}
\bibfield{author}{\bibinfo{person}{RC Weber}, \bibinfo{person}{SJ Lawrence},
  \bibinfo{person}{BA Cohen}, \bibinfo{person}{JE Bleacher},
  \bibinfo{person}{JW Boyce}, \bibinfo{person}{MR Collier}, \bibinfo{person}{D
  Draper}, \bibinfo{person}{AL Fagan}, \bibinfo{person}{CI Fassett},
  \bibinfo{person}{L Gaddis}, {et~al\mbox{.}}} \bibinfo{year}{2021}\natexlab{}.
\newblock \showarticletitle{The Artemis III Science Definition Team Report}. In
  \bibinfo{booktitle}{\emph{Lunar and Planetary Science Conference}}.
  \bibinfo{pages}{1261}.
\newblock


\bibitem[Wienrich et~al\mbox{.}(2018)]%
        {wienrich2018assessing}
\bibfield{author}{\bibinfo{person}{Carolin Wienrich}, \bibinfo{person}{Nina
  D{\"o}llinger}, \bibinfo{person}{Simon Kock}, \bibinfo{person}{Kristina
  Schindler}, {and} \bibinfo{person}{Ole Traupe}.}
  \bibinfo{year}{2018}\natexlab{}.
\newblock \showarticletitle{Assessing user experience in virtual reality--a
  comparison of different measurements}. In
  \bibinfo{booktitle}{\emph{International Conference of Design, User
  Experience, and Usability}}. Springer, \bibinfo{pages}{573--589}.
\newblock


\end{thebibliography}

\end{document}